\documentclass[journal]{IEEEtran}
\usepackage{graphicx}
\usepackage{float}
\usepackage{mathrsfs}
\usepackage{amsfonts}
\usepackage{cite}
\usepackage{color}
\usepackage{subfigure}
\usepackage{booktabs}
\usepackage{multirow}
\usepackage{algorithm}
\usepackage{algpseudocode}
\usepackage{graphics}

\usepackage{latexsym}
\usepackage{amssymb,amsbsy,verbatim,array}
\usepackage{epsfig}
\usepackage{algorithm}
\usepackage{algorithmicx}
\usepackage{algpseudocode}

\usepackage{amsmath}

\ifCLASSINFOpdf
\else
\fi
\begin{document}
	
\title{PubChain: A Decentralized Open-Access Publication Platform with Participants Incentivized by Blockchain Technology}
\author{\IEEEauthorblockN{Taotao Wang, \IEEEmembership{Member, IEEE}, Soung Chang Liew, \IEEEmembership{Fellow, IEEE}, Shengli Zhang, \IEEEmembership{Senior Member, IEEE}}
		
\thanks{T. Wang and S. Zhang are with the Guangdong Laboratory of Artificial Intelligence and Digital Economy (SZ), Shenzhen University, Shenzhen 518060, China (e-mail: ttwang@szu.edu.cn; zsl@szu.edu.cn). 
	
S. Liew is with the Department of Information Engineering, The Chinese University of Hong Kong, Hong Kong SAR, China (e-mail: soung@ie.cuhk.edu.hk)}
}	
\maketitle
	
\begin{abstract}
We design and implement Publication Chain (PubChain), a decentralized open-access publication platform built on decentralized and distributed technologies of blockchain and IPFS peer-to-peer file sharing systems. The existing publication platforms have some severe drawbacks. First, instead of promoting widespread knowledge sharing, access to publications on the platforms owned by publishers is often on a fee basis. This drawback of pay wall prevents researchers from “standing on the shoulders of giants”. Moreover, the peer review process on most all existing publication platforms (including both open-access and publisher platforms) is prone to be ineffective, since there is no proper incentive to reviewers for performing high-qualified reviews. PubChain is an alternative platform to the existing publication venues aiming to address their drawbacks. No central third-party owns the contents (i.e., papers and reviews) of PubChain. Exploiting blockchain technology, we devise an elaborate incentive scheme on PubChain to incentivize key stakeholders (i.e., authors, readers and reviewers) to participate publication activities on PubChain in a substantive manner by earning credits and rewards through self-motivated interactions. We have performed simulations to investigate the robustness of our proposed incentive scheme against fraudulent publications and reviews. We also have implemented a prototype of PubChain to demonstrate its key concepts.
\end{abstract}
	
\begin{IEEEkeywords}
Publications, Decentralization, Blockchain, Peer-to-Peer Networks, IPFS
\end{IEEEkeywords}

	\IEEEpeerreviewmaketitle

\section{Introduction}
\label{sec:introduction}
Publications of research results are an important activity to disseminate new knowledge.  ``Standing on the shoulders of giants'' is a vivid expression that points out that new discoveries and innovations are often built on prior work  by others \cite{keith2015strategic, newton2018}. Researchers thrive on free exchange of information.

\subsection{Drawbacks and Limitations of Existing Publication Platforms of Publishers}

To date, the most successful venues for academic paper publication are journals and magazines owned by large entrenched publishers, such as Nature Publishing Group, Institute of Electrical and Electronic Engineers (IEEE), Association for Computing Machinery (ACM) and Elsevier of RELX Group. These publishers publish a huge number of research papers every year. Their journals and magazines are platforms on which researchers exchange their latest research results and where latest research breakthroughs are announced. Despite their success, these publication platforms have significant drawbacks and limitations from the standpoint of key players---authors, reviewers, and readers---that matter most.

\subsubsection{Pay Wall}

The power to publish, store and share academic literature is concentrated in the hands of a few dominant publishers. These publishers are for-profit outfits. They charge authors for publications on their venues and they charge readers for accessing the papers. In other words, they charge both the producers and consumers. Furthermore, conferences organized by some publishers often charge exorbitant registration fees for conference attendance, and outrageous sums of money for page charge for pages that do not incur much additional cost on their electronic platforms. They get away with these exploitations because they can. They have built up their brands over the years. 

But who help them build and maintain their brands?  Well, they leverage the free service of editors and reviewers to maintain the quality of the publications. In most businesses, workers who do work receive compensation rather than the other way round. Publication business is an exception---publishers charge both the workers (the authors) as well as the customers (the readers) and receive free services from both the workers and the customers (the authors and readers themselves often serve as the reviewers). 

Their charges can be quite expensive to the extent that only large organizations, such as corporations, research institutions and universities, can afford the fees. The pay wall put up by the publishers excludes small organizations and individuals from accessing the latest research publications. These publishers stand in the way of knowledge dissemination and the pay wall prevents a level playing field among researchers.

\subsubsection{Information Island}

The authors are forced to transfer the copyrights of their papers to the publishers. The publishers typically do not mutually share their literature resources. This gives rise to information islands with unsynchronized contents. There are many intrinsic disadvantages associated with such isolated information islands. Readers and researchers lacking resources will have difficulty getting a complete set of past papers unless they subscribe to all these publishers. These islands are hurdles to knowledge dissemination.

\subsubsection{Disintegration of Peer Review Process}

Peer reviews of papers should be performed by experts with the same level of competence as the authors of the papers in their particular field. The peer review process is crucial to maintaining paper quality. As a rule of thumb, journals and conferences with a rigorous peer review process and with a low paper acceptance rate are considered to be more prestigious by readers and authors. 

With the growth of research participants, research papers are also growing exponentially. It is getting increasingly difficult to find quality reviewers to review the large number of papers. Competent reviewers are researchers themselves. As researchers, they need to balance their time between reviewing others’ papers and doing their own research. Unless these papers are directly related to their current research topics, they have little incentive to do the review, even if they have the technical expertise to do so. Review is a form of “technical auditing” as far as scientific papers are concerned. 

When accountants perform financial auditing for corporations and organizations, they often charge a large sum of money for their service, and as such they are obligated to do a professional job that meets a certain minimum quality threshold. Otherwise, the accountants would not receive future jobs. When reviewers perform technical auditing, reviewers receive zero compensation, and the quality of review varies much from reviewer to reviewer. There are no incentives other than the conscience of the reviewers to meet certain minimum quality target. Arguably, serious technical auditing can be a lot more time-consuming than financial auditing. Why should technical auditing be free? Are scientists worth less than accountants?

Without proper incentives, there is little reason for reviewers to spend time on paper review. As a result, because of paper explosion, many reviews are quite shallow in nature, even for prestigious venues such as IEEE. Many senior researchers (e.g., professors) may relegate the responsibility of paper review to junior novice researchers (e.g., beginning graduate students of the professors) who at least have the incentive to review papers as part of their learning process---some of them probably have no choice because their superiors ask them to do the job. Where did the money---page charges, membership fees---go? Did any go to those responsible for quality assurance? Without proper incentives to reviewers, the current peer review process can break down easily, especially in the face of paper explosion \cite{smith2006peer, adler2012new}.\footnote{ Besides the lack of incentives for reviewers, other limitations of the current peer review system are discussed in \cite{smith2006peer, adler2012new}. For example,  the review process can be slow and cumbersome; it also exhibits various forms of bias.}

\subsection{Other Publication Platforms and Services}

Literature search and citation index services, such as web of science and google scholar, can partially overcome the information island effect.  Papers from multiple publishers can be listed and their citations can be indexed. Since these services do not really publish papers, they still cannot overcome the handicaps of pay wall and peer review disintegration. 

To overcome the pay wall of publishers, Free Open Access aims to make academic literature a free public resource on a global scale. For example, the arXiv preprint system allows authors to upload their papers for free access by all researchers. By the year of 2014, more than 1 million articles have been uploaded on arXiv \cite{vence2014one}. The founder of arXiv, physicist Paul Kingsbagh, won the 2002 MacArthur award for his contribution to Free Open Access. Although Free Open Access platforms allow everyone to access research outputs freely and easily, they still suffer from peer review disintegration. In fact, arXiv does not even have a peer review process. Low-quality papers abound on Free Open Access platforms. As of today, papers published on Free Open Access platforms do not earn the same prestige that papers on the publication platforms of Publishers.

Some open access platforms such as Peerj also support peer review openly---review comments can be posted along with the published papers. However, even with this open peer review scheme, reviewers still lack incentives for investing efforts to provide high quality reviews. Therefore, the open peer review scheme does not address the drawback of peer review process's disintegration.   

All publication platforms today are centralized---they are owned or managed by a single organization. As a consequence, they are prone to single points of failure---there is no guarantee that the organization will never close the access to the database.  The power to publish, store and share academic literature is concentrated in the hands of publishers and owners of open-access platforms.

\subsection{How does PubChain Incentivize Participants}

Publication Chain (PubChain) aims to overcome the limitations of the current publication platform. PubChain is a decentralized publication platform, where authors, readers and reviewers are incentivized to participate in a meaningful and substantive manner. In particular, these key players can earn credits and rewards through self-motivated interactions. The assets of PubChain are owned by these key players, not by a separate profit-focused publisher. PubChain does not own the copyrights of the papers; the authors retain their copyrights. PubChain is not a central authority. The authors do not need permissions to publish papers on PubChain. 


In the following, we review the status quo of existing publication platforms from the standpoints of the incentives for the authors, the reviewers, and the readers.  For this purpose, IEEE is taken as a representative of publisher platforms, and arXiv is taken as a representative of Free Open Access platforms.

Incentives for Authors:

\begin{enumerate}
	\item	Visibility - The most important motivation for authors is that their papers are downloaded and read by many. This is successfully achieved by IEEE already. It is also achieved to some extent by arXiv given its open access nature.
	\item	Prestige and Recognition by Peers - Well written papers with good results are recognized by peers. This is successfully achieved by IEEE already. As of today, the quality of papers in arXiv varies widely because of the lack of a review process. Having an arXiv paper by itself does not earn recognition from  peers. 
	\item	Time Stamping - Claiming the first to do something. This is achieved by IEEE to some extent; however the time stamps are not immediate. Time stamping is more immediate with arXiv.
	\item	Low Cost -  Publishing on IEEE venues is very costly. Uploading papers to arXiv incur no cost. 
	\item	Continuous Improvement of Publications - Authors can submit revised versions of papers based on  feedback and reviews on the platform. If this can be achieved, research publications, like software, will have a life of its own in that it can be continuously improved. Papers published in IEEE go through a few reviewers only. And once accepted and published, the publications are a static record. arXiv allows authors to submit new versions of the same paper. However, there is a lack of feedback by reviewers that add quality to the new versions. 
	\item	Financial Incentive - To most authors, making money from publications probably ranks low as an incentive.  That said, as far as we know, IEEE (and other publishers) does not pay authors of significant papers that add prestige to their journals and magazines. Occasionally, prize paper awards come with only a small token amount of money as a goodwill gesture.  There are no financial incentive schemes on open-access platforms either.   
	
\end{enumerate}

Incentives for Reviewers:

\begin{enumerate}
	\item Reward - Good reviews should be rewarded financially or rewarded by other means. Paper review is an “auditing” process. Why should the efforts of paper reviewers be free especially if most reviewers do not gain recognition from the efforts? Technical people have been exploited to a large extent in that regard. IEEE certainly does not provide strong incentives for reviewers to do a good job. Reviewers are not participants on the arXiv platform. 
\end{enumerate}

Incentives for Readers:
\begin{enumerate}
	\item	Good and Relevant Papers - Readers, who are often researchers themselves, want to find good and relevant papers quickly. This is achieved by IEEE and arXiv. 
	\item	Interactions with Authors and Reviewers - Readers can obtain answers from the authors directly on the platform. Each paper could also have a FAQ managed by the author, but with contributions from the other readers and reviewers. These are very little open interactions and debates between readers, authors, and reviewers, on  IEEE and arXiv.
	
\end{enumerate}

PubChain is a blockchain-based publication platform designed to achieve the above incentives for all stakeholders through an incentivized ecosystem. We next discuss why we choose blockchain as our solution.

\begin{table*}
	\centering\centering
	\caption{Comparison of different publication platforms}
	\label{table}
	\begin{tabular} {|c|c|c|c|c|}
		\hline
		& 
		Pay Wall&    Single Point of Failure & Disintegration of Peer Review Process\\
		\hline
		Publishers & $\times $ &    $\times $  &  $\times $  \\
		\hline
		Free Open Access & $\surd$   &   $\times $   &  $\times $   \\
		\hline
		Consortium Web App & $\surd$   &      $\surd$  &  $\times $ \\
		\hline
		PubChain &   $\surd$ &    $\surd$  &   $\surd$  \\
		\hline
	\end{tabular}
	\\
	\footnotesize{$\times $: having such problem. \\ $\surd$: having no such problem.}
	\label{tab1}
\end{table*}

\subsection{The Motivation and Justification for the use of Blockchain}
	
Blockchain is a decentralized and distributed digital ledger that stores data in chronological order in a way that can ensure that the data in the chain cannot be falsified \cite{puthal2018everything, christidis2016blockchains, consensus2019}. It is natural to employ blockchain to solve the problems of pay wall and single point of failure. Although other solutions (e.g., a web application run by a consortium) can also address some problems of the existing publication platforms, i.e., pay wall (if the consortium's platform subscribes to free open access), single point of failure, there are still a number of advantages enjoyed by blockchain that cannot achieved by these solutions.  

Using blockchain, our PubChain can construct a decentralized publication platform that does not belong to any entities. Besides solving the problems of pay wall and single point of failure, the decentralized publication platform can establish an ecosystem for publication activities, where authors, reviewer and readers are incentivized to make positive contributions.

We believe that there is no free lunch---but good lunch should not cost too much, and people who contribute to the lunch quality should be rewarded. For certain production and sales activities, there must be people who pay money (the ones who enjoy the services/products) and people who receive the money (the ones who provide the services/products). Authors need to pay for their publications, due to that they use others' services (i.e., platforms' publication services, reviewers' review services). However, on all current publishers' platforms (i.e., conferences, journals), authors just pay the publishers and not the reviewers, and fees paid to publishers are rather high. On free open-access platforms, authors do not pay anybody and there is no review service provided. Besides the platforms' publication services, reviewers' services are also important to paper quality assurance. Reviewers provide some sort of a "consumer reports" service. However, reviewers on current publication platform lack incentives to perform high quality reviews. Therefore, we design PubChain to incentivize reviewers to provide high quality reviews by letting them to receive an amount of financial rewards commensurate with the quality of their reviews.

How to credit reviewers is not an easy task, since how to evaluate their review comments and scores is not straightforward. Using blockchain technology, we design an incentive mechanism for authors, reviewers and readers that incentivizes these participants to form an ecology that promotes paper quality and achieves free open access at the same time (see Section IV). In particular, we devise a decentralized scoring mechanism that is robust to dishonest scoring behaviors (see Section V). We compare the current existing publication platforms, a web application run by a consortium and our PubChain by indicating the problems that can be solved by them in TABLE I.

\subsection{Related Work and Our Contributions} 
Refs. \cite{smith2006peer, adler2012new} discussed the limitations of the current publication platforms, especially the limitations imposed by the peer review process. Works \cite{spearpoint2017proposed, janowicz2018prospects} proposed to incentive reviewers using digital cryptocurrency of blockchain. Work \cite{sharples2016blockchain} proposed to permanently record educational records (i.e., exam credentials, record of learning, the authorship of something) and the corresponding reputation rewards on blockchain. Works \cite{NizamuddinHS18} and \cite{tenorio2019towards} proposed to use blockchain to record the copyright of publications and use peer-to-peer file storage system IPFS to store the publications. 

These existing works mainly talked about the idea of using blockchain to record the copyright of papers and/or issuing tokens to authors/reviewers. However, a complete framework with rigorous designs is still lacking. Our PubChain is a complete framework with rigorous designs for implementing blockchain based publication platform. Compared to these existing works, PubChain has the following new contributions.

\begin{itemize}
	
	\item First, PubChain has an incentive mechanism to authors, reviewers and readers that incentivizes these participants to form a self-sustaining ecology that promotes paper quality on an open access platform (Section IV). 
	
	\item Moreover, PubChain has a decentralized scoring mechanism that is robust to dishonest scoring behaviors (Section V). 
	
	\item Finally, we consider many practical aspects of the system, including the system architecture (Section III), and the financial model (Section II).

\end{itemize}

\section{Solution of PubChain}

\subsection{Design Concept}

\begin{figure*}[!t]
	\centering
	\includegraphics[width=5.5in]{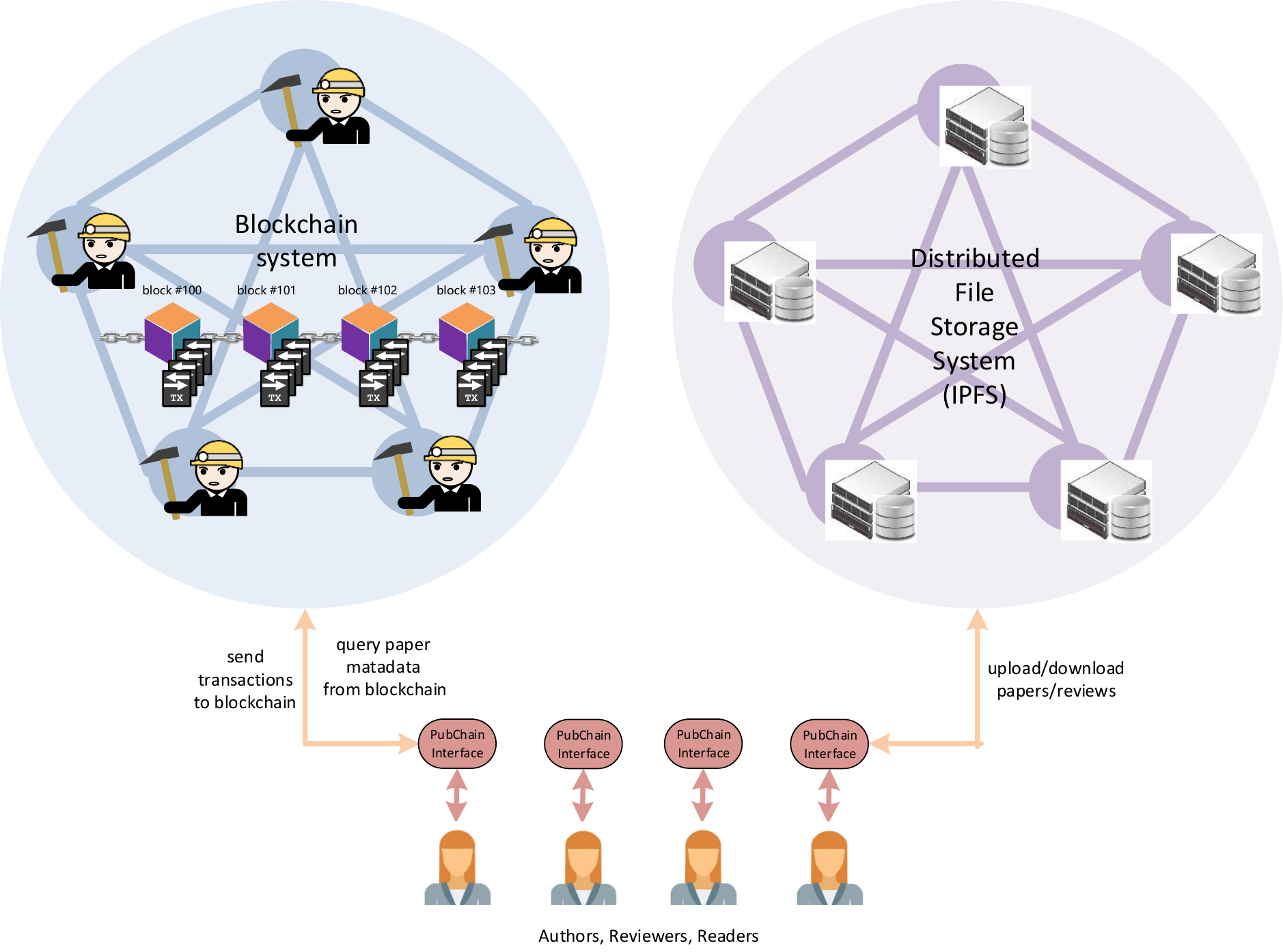}
	\caption{The overview of PubChain platform, where we have three entities: a group of publication players (authors, reviewers, readers), a blockchain system sustained by miners, an IPFS system with distributed storage nodes.}\label{fig1}
\end{figure*}

A central design concept of PubChain is to use blockchain \cite{puthal2018everything, christidis2016blockchains, consensus2019} and off-chain peer-to-peer distributed file storage (i.e., InterPlanetary File System (IPFS) \cite{benet2014ipfs,wang2018blockchain, chen2017improved}) as building blocks to decentralize the publication platform. Such decentralization also means that there is no single central party that controls the running of the platform. If properly designed, the decentralized system can also be more robust than a centralized system given its replication of data across multiple parties. 

PubChain uses the IPFS system as the database system for storing papers. IPFS is a distributed and decentralized storage system consisting of a network of peer-to-peer nodes. The techniques and features of IPFS can be found in  \cite{benet2014ipfs}. With IPFS, papers are content-addressed in the database. Authors can back up their papers to the network and freely download papers without the risk of single-point failure. The IPFS repository is physically owned by all users and not by a single entity.

PubChain exploits blockchain technology to confirm the registration of the paper ownership, to track index citations, and to incentivize participants. Blockchain is a distributed and decentralized append-only ledger for digital assets. Data in blockchain is replicated and shared among all the participants. Past records are made tamper-resistant through its append-only paradigm.  There are many successful existing blockchain systems, e.g., Bitcoin \cite{nakamoto2008bitcoin}, Ethereum \cite{buterin2013ethereum}. We can reuse and modify their software code to build the blockchain of PubChain.  

The operation of PubChain blockchain is divided into two consecutive phases, with the first phase being a temporary phase before the final second phase takes over. In the first phase, PubChain operates as a consortium blockchain using the Proof-of-authority (PoA) consensus protocol \cite{parity}. In the second phase, PubChain operates as a public blockchain using the Proof-of-work (PoW) consensus protocol \cite{nakamoto2008bitcoin}.

Fig. 1 gives an overview of the PubChain platform. There are three entities in the platform: a group of publication players, a blockchain system sustained by miners, an IPFS system with distributed storage nodes. The blockchain system and the IPFS system are the infrastructure of PubChain. A network node can be a miner of blockchain or/and a storage node of IPFS. Blockchain miners run the distributed consensus protocol to maintain the data on blockchain.  In the consortium blockchain phase, the miners are the super nodes that are selected to run the PoA protocol. In the public blockchain phase, the miners are the nodes that devote their computing powers to solving hash puzzles of the blockchain. IPFS storage nodes share their memory space for the distributed and persistent storage for PubChain.  Through a PubChain interface, the publication players (i.e., authors, reviewers, and readers) interact with the blockchain and IPFS systems in the conduct of their activities on PubChain. We have developed a PubChain system that combines blockchain and IPFS. We will describe the system architecture of PubChain in Section III. 

When an author uploads his/her paper to PubChain, the paper is time-stamped and registered on Pubchain as a permanent record.  The citation index for every paper is also tracked on PubChain. Tokens are used to financially incentivize players to engage in publication activities on PubChain and to incentivize miners to sustain and maintain PubChain. We will elaborate our proposed incentive mechanism in Section IV.

The tokens issued by PubChain are called PubCoins. PubChain is a non-profit project and we will not sell the issued PubCoins through initial coin offering (ICO) and private placements to any other entity to make money. PubCoins will be distributed to all the participants as the rewards for their contributions to the platform, rather than to the project team or other organizations.  

To endow PubCoin with real monetary value, we design PubChain as a side chain of another parent chain whose tokens are in wide circulation and are considered to have real monetary values, e.g., Bitcoin, Etherum, Bitcoin Cash. Using the two-way pegging technique of side chain \cite{back2014enabling}, we can transfer the tokens on the parent chain to PubChain and vice versa. This concept is illustrated in Fig. 2. The technical details of two-way pegging and side chain can be found in \cite{back2014enabling}. At the beginning stage, PubChain operates separately from the parent chain, and PubCoin has no real monetary values. Donation to PubChain can be injected into PubChain from a parent chain using two-way pegging, and PubChain will then operate as a side chain after that. We discuss the details about the financial model of PubChain in the Section II.B.

\subsection{Financial Model}

With the crypto-currencies provided by blockchain systems, PubChain aims to establish the following financial model for the world of publications.  

A certain amount of PubChain tokens (PubCoins) are issued to the participants on PubChain. Corresponding to the two phases of blockchain systems, the establishment of the value over PubCoins is also divided into the following two phases:

\begin{itemize}
	\item Phase I (Consortium blockchain phase): In this phase, as a bootstrap incentive scheme, a certain amount of new PubCoins is issued to each PubChain user when he/she first registers  as a user. To endow PubCoins with real monetary values, we adopt the two-way pegging technique of side chain to transfer the values of other cryptocurrencies (that already have real prices on the market) to PubChian.  On one parent chain, we lock a certain amount of the cryptocurrency tokens to a special address and we also send the simplified payment verification (SPV) \cite{nakamoto2008bitcoin} proof of this token-locked transaction to PubChain. The cryptocurrency tokens owned by the special address cannot be transferred to other address by spending: these cryptocurrency tokens are simply a “reserve” to endow PubChain tokens with real monetary values.  On PubChain, the miners will package the transaction sent from the parent chain into a block for broadcast to the whole Pubchain network. Then, a block on PubChain issues a number of PubCoins to the users of PubChain (with two-way pegging, these issued PubCoins can be sent to the parent chain to unlock the locked cryptocurrencies on the parent chain).  In this manner, the issued PubCoins are linked to the locked cryptocurrencies on the parent chains; the value of the PubCoins are endorsed and determined by the total amount of the locked tokens.
	
	\item Phase II (Public chain phase): In this phase, a certain amount of PubCoins are issued in each block. These issued tokens will be given as rewards to the miner as well as  to the authors and reviewers that contribute to PubChain.  How the rewards are distributed among the players will be explained later. The amount of the rewarded PubCoins in each block is constant and does not vary from block to block. This means that the total amount of tokens issued increases over time and is unlimited. No other cryptocurrency is transferred to PubChain anymore in Phase II. PubChain is operated as a decentralized central bank that constantly issues new tokens to adapt to the expansion of the whole economy on the platform.
	
\end{itemize}

\begin{figure}[!t]
	\centering
	\includegraphics[width=3.4in]{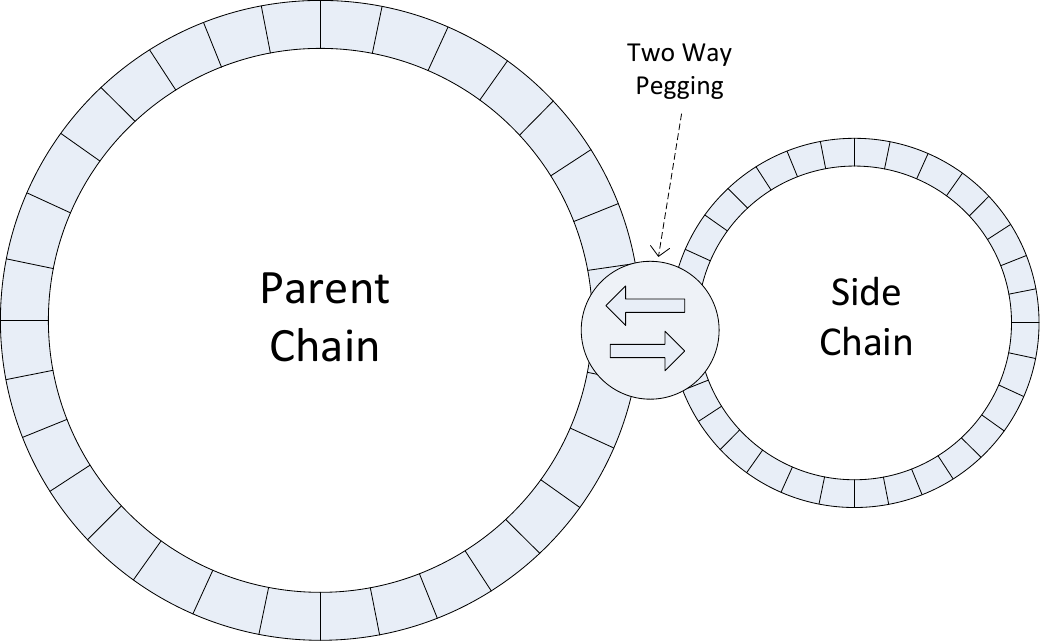}
	\caption{Illustration of the concept of side chain and its two-way pegging.}
\end{figure}

Fig. 3 illustrates the above two-phase financial model of the PubChain system. A few remarks are as follows: 
\begin{itemize}
	\item There is a important difference between the cryptocurrency endorsement mechanism in Phase I of  PubCoin and the ICO activities of other projects. Using the two-way pegging technique, Pubchain cryptocurrency endorsement can guarantee the cryptocurrencies transferred to PubChain belong to the entire PubChain network rather than to a single entity (i.e., not controlled by one entity); nobody can embezzle these locked cryptocurrencies and spend them. For ICO, there is a high risk that the institution or individual controlling the project will abscond with the raised cryptocurrencies. 
	
	\item Phase I is similar to the Bretton Woods system (the system for monetary and exchange rate management established in 1944 \cite{van1978bretton}) where the US dollar is linked to gold (all involved countries confirmed the official price of 35 US dollars per ounce of gold in January 1944) and the currencies of other countries maintain fixed exchange rates with the US dollar. Under Bretton Woods system, the credit and the value of the US dollar is supported by gold. 
	
	\item The purpose of Phase I is to inject real monetary values into PubCoins to incentivize participants to conduct publication activities over PubChain.  This phase is very important for cold-booting PubChain. As more active participants join PubChain and as the value of PubChain to publication players are demonstrated, more participants can be brought in, whether they are incentivized by money or by the value of PubChain as a publication venue. At some point, a vibrant ecosystem will be established, and there will be no need to inject the values of other crypto-currencies into PubChain. PubChain then operates normally with its financial value tied to its use value: PubChain then enters Phase II of its operating model.
	
	\item In Phase II, the amount of the rewarded PubCoins in every block keeps constant all the time, which means the total amount of tokens issued increases over time and is unlimited. This is different from the token issuing mechanism of Bitcoin that limits its total amount of tokens.  This implies the underlying financial models of PubCoin and Bitcoin are different. Bitcoin is more like gold, and its value comes from its scarcity; PubCoin is more like a currency, and the issuer continues to issue this currency to adapt to the gradual expansion of the entire economy: if more work is done (in this case, more publications, more reviews, more readers participations), more currency will be issued to support the increased scale of the economy. Phase II is similar to the current US dollar system after the collapse of the Bretton Woods system since the early 1970s.  
	
\end{itemize}

\begin{figure}[!t]
	\centering
	\includegraphics[width=3.6in]{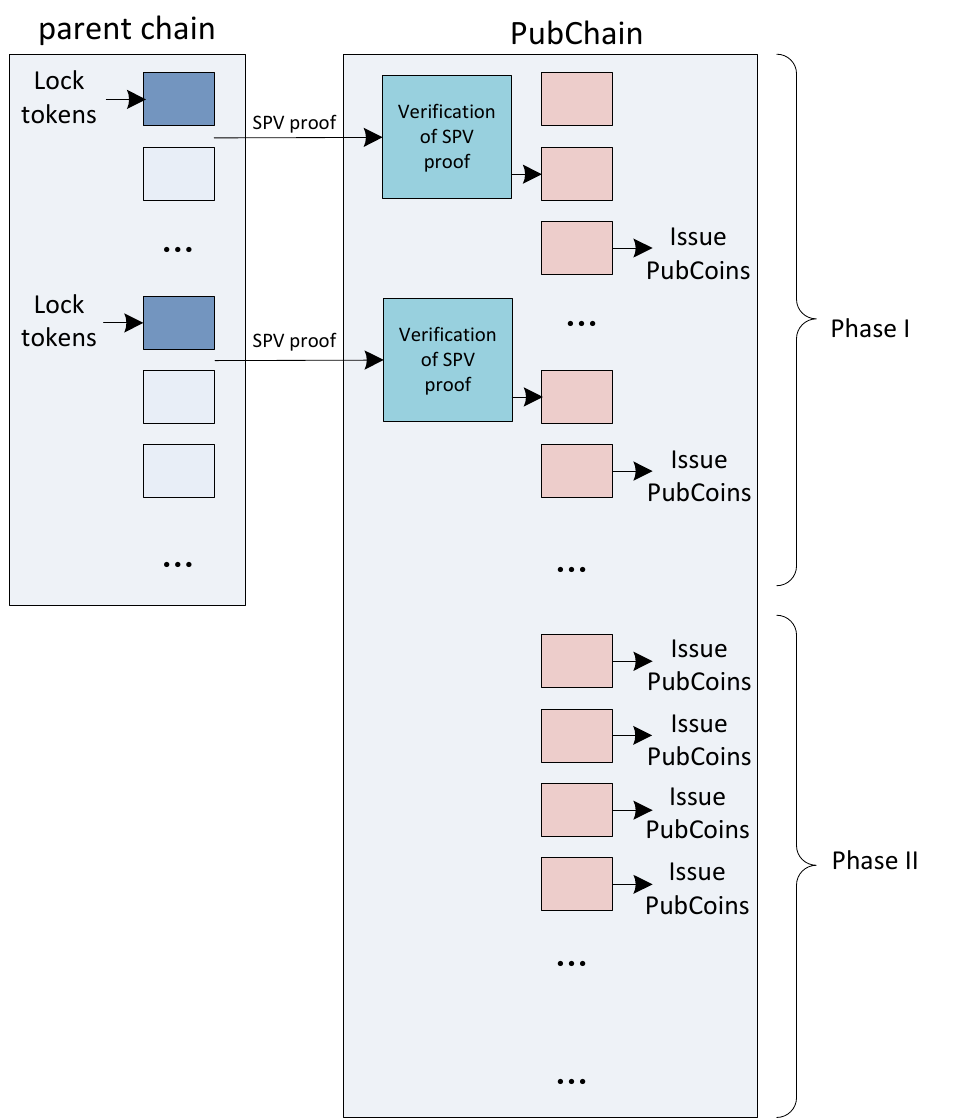}
	\caption{The illustration of the two-phase financial model of the PubChain system.}
\end{figure}

\section{System Architecture of PubChain}

Recently, blockchain-based decentralized data storing and sharing networks were investigated in \cite{zyskind2015decentralizing, azaria2016medrec, wu2018tsar, shafagh2017towards, liu2017blockchain}.  PubChain has many technological similarities with these existing blockchain-based decentralized data systems. One major difference of PubChain, however, is that the data stored and shared on PuhChain are very specific, i.e., papers. This difference adds a set of special technical requirements to the design and implementation of PubChain. 

PubChain has the following technical requirements:
1) Any node should be able to upload and share papers on the platform.
2) The platform should provide scalable data transmission and storage capabilities.
3) Nodes should be incentivized to upload papers to PubChain to derive benefits from the uploaded papers.
4) The platform should be able to identify and evaluate the quality of papers.
5) The platform does not keep ownership of papers. 

In order to realize these technical requirements, PubChain is built on a completely decentralized system architecture. As shown in Fig. 4,  the system architecture of PubChain consists of four layers: blockchain layer, virtual machine layer, routing layer and storage layer. Following the design principle proposed in \cite{ali2016blockstack}, this architecture decouples the control plane (that consists of the blockchain and virtual machine layers) and the data plane (consisting of the Routing and storage layers). We describe the functions of the four layers in the following.

\subsection{Blockchain Layer} 
The bottom layer is the blockchain layer. The Bitcoin-like blockchain system is a tamper-proof distributed ledger that is suitable for recording small data but not suitable for processing big data. For the design of PubChain, the blockchain systems are exploited to record the metadata of papers; the blockchain systems also record the operation commands sent by the nodes and enable consensus on the execution order of operations.  In a nutshell, the blockchain layer realizes a global state recorder for the PubChain platform.   

\subsection{Virtual Machine Layer}  
Above the blockchain layer is the virtual machine layer. The API functionalities provided by the script languages on the Bitcoin-like blockchain systems are very limited. For example, the Bitcoin blockchain can only perform simple operations such as issuing and recording transactions. For PubChain, the logic functionalities reside in the more versatile virtual machine layer. We can define new operations in the virtual machine layer without changing the underlying blockchain layer. The virtual machine layer reads the recorded metadata of papers and the operation commands from the blockchain layer and executes these operations accordingly.  

Our current reference implementation of PubChain uses the Ethereum Virtual Machine (EVM) smart contract mechanism \cite{buterin2013ethereum} in its virtual machine layer. Since the EVM smart contract is Turing complete, we can realize the functionalities required by the PubChain platform and easily incorporate other new functionalities as they arise. Another choice for the virtual machine layer is the virtual chain mechanism proposed in \cite{ali2016blockstack}. We can also  realize the logic of PubChain by predefining the required operations using the technology of  virtual chain.  Although virtual chain is not Turing complete,  it is a lightweight design whose advantages include better reliability, security and performance.

\subsection{Routing Layers}  
Above the virtual machine layer is the routing layer. The main function of the routing layer is to allow the virtual machine layer to obtain the addresses of papers in the storage system. PubChain separates routing requests (i.e., how to locate papers) from the actual storage of papers. This avoids the need for PubChain to adopt a specific storage backend, allowing the coexistence of multiple forms of storage backends, including centralized database, commercial cloud service, and peer-to-peer distributed file sharing systems (e.g., IPFS).

In the implementation of PubChain, a subset of publication player nodes, blockchain miner nodes, IPFS storage nodes forms a peer-to-peer network based on distributed hash table (DHT) \cite{stoica2001chord} for storing the routing files. When the virtual machine layer sends an address resolution request to the routing layer, the routing layer first looks for the corresponding routing file according to the target hash value of the paper sent by the virtual machine layer through the DHT mechanism. With the routing file, we can then get the URL of the specific storage locations (such as cloud service: https://, IPFS storage: ipfs://)\footnote{One paper can be stored in multiple storage systems to ensure its availability. The IPFS system is always used to store papers for ensuring free open access. If the URL of a cloud service is used, the access to the paper is via a central server. If the URL of IPFS is used, the access to the paper is via the DHT embedded in IPFS system.}. A data request is then initiated to the corresponding storage system via the URL.

\subsection{Storage Layer}   
The top level of the PubChain is the storage layer that stores the actual data of papers. Each stored paper is signed by the owner's private key to claim its ownership. Because papers are stored outside of the blockchain, PubChain can support the storage of papers with any size and can use a variety of storage backends. Nodes do not need to trust the storage systems because they can verify the integrity of downloaded paper in the virtual machine layer.  Multiple forms of data storage can be mounted to the storage layer. For example, some large institutions (such as universities, companies) can set up their own centralized databases to back up the papers on PubChain for their own use. However, the decentralized peer-to-peer IPFS storage system is always included to ensure the accessibility of the papers. 

\begin{figure}[!t]
	\centering
	\includegraphics[width=3.4in]{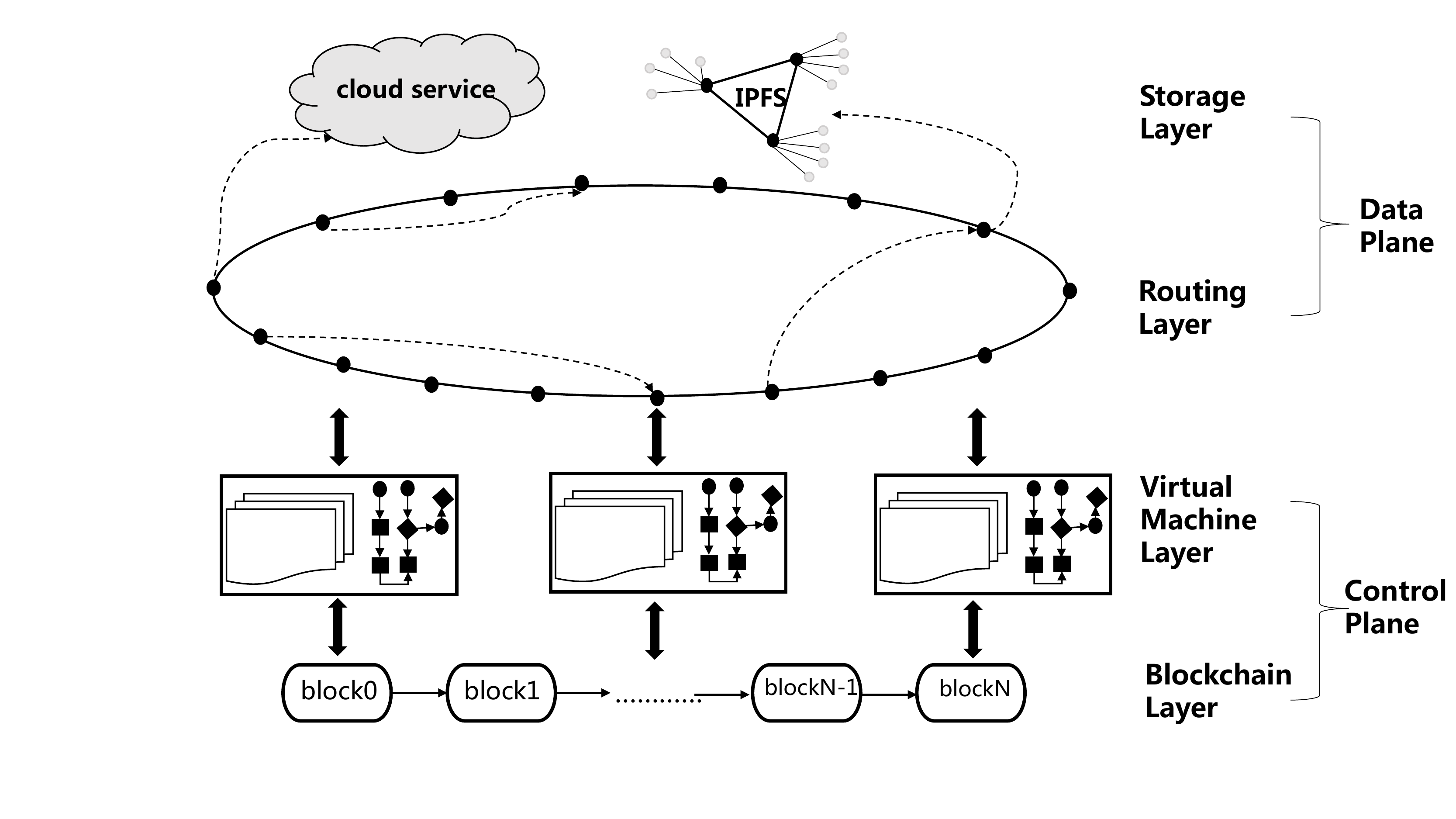}
	\caption{The System Architecture of PubChain.}
\end{figure}

%
%

\section{Incentive Mechanisms of PubChain}

Papers on PubChain will attract large readership only if it is a reputable publication platform. Authors and reviewers contribute toward making PubChain a quality publication platform by submitting high-quality papers and reviews. In that light, effective incentive mechanisms to encourage substantive and meaningful participation by authors and reviewers are a core part of Pubchain.  In this section, we describe the incentive mechanisms of PubChain\footnote{We focus  on the incentive mechanism to publication players here. The incentives to blockchain miners are the mining rewards and transactions fees. Filecoins \cite{benet2018filecoin} are the incentive to IPFS storage nodes who share their storage spaces.}.

\subsection{Incentive Mechanism to Authors }

In Pubchain, an author submits her/his paper via a transaction ${T_{post}}$. The transaction ${T_{post}}$  includes meta-information associated with the submitted paper, such as the author’s address on the blockchain, the paper’s IPFS address, title, keywords, and the transaction hashes of papers cited by the paper.

The author pays $X$ PubCoins in the transaction of each submitted paper.\footnote{On PubChain, when users generate the blockchain addresses, they must provide their affiliation emails or ORCID IDs, and each affiliation email/ORCID ID can only be linked to one address. At the beginning stage of PubChain, we will send a few tokens to each address for free, so that authors can upload their papers to the platform. After PubChain is widely accepted by authors and it has gained recognition as a reputable publications platform, we will stop issuing free tokens. From then on, to publish papers, authors will have to find ways to earn tokens (e.g., by serving as reviewers for others’ papers) or else purchase tokens from others to pay the author fees. We believe the payment scheme will not reduce the number of submitted papers.} A fraction of ${a_1}X$ tokens, $0 < {a_1} < 1$, are allocated to a\emph{reviewer bonus pool}. The tokens in the reviewer bonus pool are distributed to reviewers according to the mechanism presented in Section IV.B. Another fraction of  ${a_2}X$ tokens,  $0 < {a_2} < 1$, ${a_1} + {a_2} < 1$  are given to the papers cited by the submitted paper \footnote{Only authors of cited papers having a registered account with PubChain will be rewarded. Each Paper is identified by the hash of the transaction that registered the paper on the blockchain.}. The remaining  $\left( {1 - {a_1} - {a_2}} \right)X$ tokens are taken as the transaction fees given to the miner that records the transaction onto blockchain.

In the public-chain phase of PubChain, every mined block contains a coinbase transaction that mints $Y$ new PubCoin tokens. Among these minted tokens, a fraction of ${b_1}Y$  tokens, $0 < {b_1} < 1$, are released to the reviewer bonus pool, a fraction of ${b_2}Y$ tokens, $0 < {b_2} < 1$,  ${b_1} + {b_2} < 1$ are released to authors as rewards according to the following reward distribution mechanism, and the remaining $\left( {1 - {b_1} - {b_2}} \right)Y$  tokens are released to the miner of that block.

To incentivize authors to submit quality papers, rewards are distributed to authors according to the review scores of their papers. Specifically, when a new block is mined, the author of paper $i$  receives a reward of $G_i$ PubCoins from the new block, computed as follows: 
\begin{equation}\label{1}
{G_i} = {b_2}Y \times \frac{{\max \left( {{S_i} - \lambda ,0} \right)}}{{\sum\limits_i {\max \left( {{S_i} - \lambda ,0} \right)} }}
\end{equation}
where ${S_i}$ is the current review score of paper $i$  (computation of the current review score ${S_i}$  of paper $i$  will be presented in Section V.A),  $\lambda $  is a quality threshold for papers, the summation of $i$  is over all papers that were been published on PubChain during the previous $M$  block intervals. In other words, a paper will be rewarded within a reward window of $M$ blocks after it is published on PubChain.

The review score ${S_i}$ of paper $i$ is initialized to zero, ${S_i} = 0$, and thus $\max \left( {{S_i} - \lambda } \right.$ $\left. {,0} \right) = 0$ initially. With the threshold $\lambda $, posting and overwhelming PubChain with low-quality papers whose review score does not pick up over time will not be rewarding. Furthermore, since $X$ PubCoins are charged for each posted paper, there is a disincentive for authors to submit low-quality papers.

Besides receiving rewards from good reviews, a paper can also receive rewards when it is cited by another paper.  Specifically if paper $j$ cites $K$ other papers that are also posted on PubChain,  ${a_2}X$ tokens paid by  paper $j$ will be given to the authors of the $K$ cited papers. Each cited paper $i$ will receive ${{{a_2}X} \mathord{\left/
		{\vphantom {{{a_2}X} K}} \right.
		\kern-\nulldelimiterspace} K}$ tokens from paper $j$.  In this way, if a paper has a long lasting influence on other papers, it may continue to receive rewards through the citation mechanism (i.e., long after the review reward window has transpired, citation reward may continue).

\subsection{Incentive Mechanism to Reviewers}

In Pubchain, a reviewer submits the review of a paper by sending a transaction $T_{review}$ to the  blockchain. The transaction  $T_{review}$ includes the hash of the transaction that publishes the paper, the  numerical score of the paper given by the reviewer, and the blockchain address of the reviewer. In addition to the numerical score, reviewers can also write comments on papers. Insightful comments are useful to the authors in terms of improving future versions of their papers; they also let readers identify high-quality papers. The comments on papers are stored on IPFS and their IPFS address  are included in the transaction $T_{review}$ that is sent to PubChain for record keeping. 

\begin{figure}[!t]
	\centering
	\includegraphics[width=3.5in]{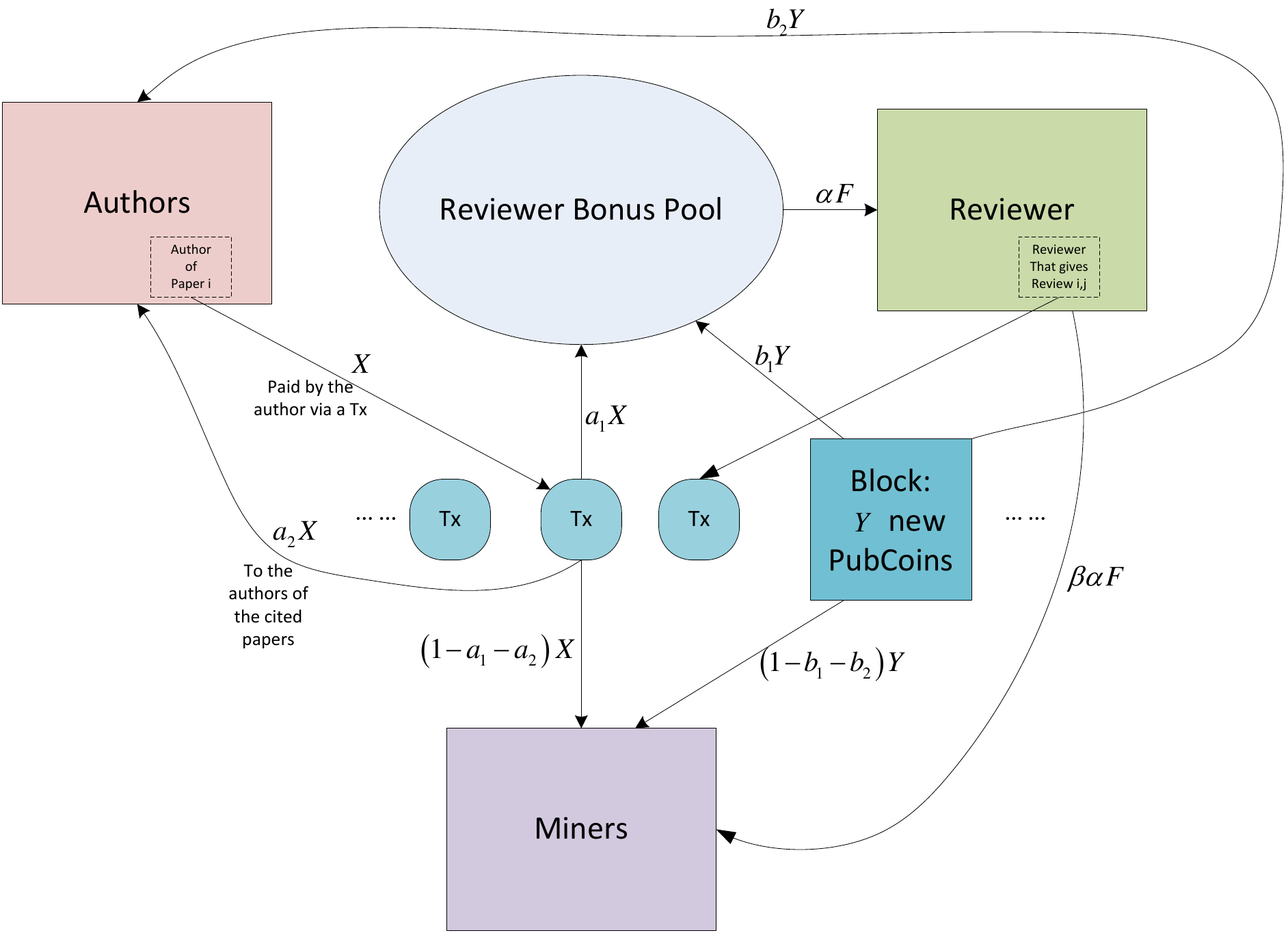}
	\caption{Token flows of the incentive mechanism in PubChain.}
\end{figure}

We denote the review of paper $i$ by reviewer $j$ by ${R_{i,j}} \buildrel \Delta \over = \left( {{Z_{i,j}},{C_{i,j}}} \right)$ where ${Z_{i,j}}$ is the numerical score and ${C_{i,j}}$ is the comments. PubChain treats the comments by reviewers as some sort of a “special paper” that are reviewed by readers –  paper reviews are also reviewed, but with a numerical score only. The score of review $j$ depends on its review numerical scores given by readers. Readers will not give high scores to a paper review with only a numerical score without insightful comments. Review $j$ of paper $i$ receives a reward of ${g_{i,j}}$ PubCoins computed as follows: 
\begin{equation}\label{2}
{g_{ij}} = \alpha F\frac{{\max \left( {{W_{i,j}} - \lambda ,0} \right)}}{{\sum\limits_{i,j} {\max \left( {{W_{i,j}} - \lambda ,0} \right)} }}
\end{equation}
where ${W_{i,j}}$ is the current average score of comment ${C_{i,j}}$,  $F$ is the total reward in the review bonus pool during the current block interval \footnote{In the consortium-chain phase of PubChain, $F = \left( {1 - \alpha } \right)F' + {a_1}XN$, where $F'$ is the total reward in the review bonus pool during the last block interval, $N$ is the number of the published papers and  ${a_1}XN$ is the total amount of the tokens paid by the published papers during this block interval. In the public-chain phase of PubChain, $F = \left( {1 - \alpha } \right)F' + {a_1}XN + {b_1}Y$, where ${b_1}Y$ is the amount of tokens released to the review bonus pool by the current block interval.}, $\alpha $ is a ratio ($0 < \alpha  < 1$) that governs how much bonus in the current pool are distributed to reviewers during this block interval, the summation of $i$ and $j$ is over the comments recorded onto PubChain during the previous $M$ blocks. The bonus not used in the current block, $(1 - \alpha )F$, will be kept in the pool for release in subsequent blocks.

To incentive miners to include review transactions into their blocks, a fraction $\beta  \ll 1$ of the reward obtained by a review ${g_{ij}}$ (i.e. $\beta {g_{ij}}$ tokens) is released to the miner who included this review transaction into its block.  Therefore, during each block interval, $\beta \alpha F = \beta \sum\nolimits_{i,j} {{g_{ij}}} $ tokens from the review bonus pool are released to miners who included review transactions associated with all reviews $j$ in the past $M$ blocks.

Fig. 5 illustrates the token flows associated with our incentive mechanism. The incentive mechanism relies on the scores that can objectively reflect the qualities of papers and reviews. The next section will present a decentralized scoring system that can prevent malicious nodes from tampering with scores.

\section{Decentralized Scoring System of PubChain}

The financial rewards of PubChain are issued to authors and reviewers according to the scores of their papers and their reviews. To earn more rewards, malicious nodes may deliberately give scores that deviate from the true quality of papers and reviews. Therefore, a decentralized scoring system that can ensure objective scores in the presence of malicious nodes is very important.  In this section, we first propose a decentralized scoring system to compute the scores of papers and reviews. We then perform simulations to investigate the integrity of the proposed decentralized scoring system.

\subsection{Decentralized Scoring System}
The effective score ${W_{i,j}}$ of review ${R_{i,j}} = \left( {{Z_{i,j}},{C_{i,j}}} \right)$ is computed by averaging readers' scores on ${R_{i,j}}$. If review ${R_{i,j}}$ has received scores from less than ${N_{rs}}$ readers, its effective score is fixed to  ${W_{i,j}} = 0$;  otherwise,  the effective score ${W_{i,j}}$  of review ${R_{i,j}}$ is obtained by excluding the highest and lowest 10\% readers' scores and then averaging the remaining scores. To avoid conflict of interest, if a participant has submitted a review of a paper, she/he cannot score the other reviews of the same paper as a reader. In addition, to avoid score flooding, a reader can at most score ${N_{rc}}$ review comments of the same paper.\footnote{For implementation, we need a way to identify participants on PubChain and associate each participant ID with a unique address on blockchain. To achieve this, we can use the affiliation emails or ORCID IDs of the participants as their IDs on PubChain. Moreover, identifying reviewers using their affiliation emails or ORCID IDs means that our scoring system is a real-name system. This can avoid the dishonest review behavior in which a paper is intentionally given high scores by many reviewers from the same affiliation.} The effective score ${W_{i,j}}$ of review ${R_{i,j}}$ is for two purposes. First, it is used in (2) to incentivize reviewers to perform high-quality reviews. Second, it is used to compute the effective score ${S_i}$  of  paper $i$.  

We employ the review results of paper $i$,  encoded in the form of ${\left\{ {\left( {{Z_{i,j}},{C_{i,j}},{W_{i,j}}} \right)} \right\}_{j = 1,2, \cdots }}$  to compute the review score ${S_i}$ of paper $i$. First, we normalize the scores ${W_{i,j}}$ of review ${R_{i,j}}$ given by the readers as: 
\begin{equation}\label{3}
{\widetilde W_{i,j}} = {{{W_{i,j}}} \over {\sum\limits_j {{W_{i,j}}} }}
\end{equation}
for all $j$. The normalized score ${\widetilde W_{i,j}}$ takes value between 0 and 1.  Then, we compute ${S_i}$ as a weighted sum of scores ${Z_{i,j}}$ given by reviewers to paper $i$ using the normalized scores  ${\widetilde W_{i,j}}$ as their weights:  
\begin{equation}\label{4}
{S_i} = \sum\limits_j {{{\widetilde W}_{i,j}}{Z_{i,j}}}
\end{equation}
The computed ${S_i}$ is an evaluation on the quality of paper $i$ and is used to reward the author by the reward distribution mechanism. In essence, the effective score ${W_{i,j}}$  made by readers to review ${R_{i,j}}$ reflects the quality of that review and is an indication of the extent to which readers agree with the score  ${Z_{i,j}}$  by reviewer $j$ on paper $i$.

\subsection{Simulation Investigations}

We next present simulation results to validate that our proposed decentralized scoring system can ensure fair reviews of papers, even in the presence of adversary reviewers with a biased interest. 

Consider one poor-quality paper, paper $i$ with a ground-truth score of $S$.  The author of this paper is an attacker who wants to gain more rewards by controlling a set of malicious nodes faking as reviewers and readers so that the paper can obtain a much higher score $S' \gg S$ on the platform. We assume the scores ${\{ {Z_{i,j}}\} _j}$ for a paper $i$ given by honest reviewers are Gaussian distributed with mean $S$ and variance $\sigma _s^2$.  The scores given by honest readers to a particular review $j$ of paper $i$  are Gaussian distributed with mean ${W_P} - \left| {{Z_{i,j}} - S} \right|$ and variance $\sigma _s^2$, where ${W_P}$ is the mean score for a “perfect review” that assigns the same score to the ground-truth score (i.e., ${Z_{i,j}} = S$).

We consider two strategies for the attacker. The first strategy is to have all malicious nodes serve as reviewers of the paper. All malicious nodes will give a high review score  ${Z_{i,j}} = S'$, for all $j \in MS$, where $j \in MS$ is the set of the malicious nodes controlled by the attacker. In our simulations, we assume there are totally 1000 review scores given by reviewers to paper $i$, among which ${N_{mn}}$ scores are given by the malicious nodes. We assume each review ${R_{i,j}} = \left( {{Z_{i,j}},{C_{i,j}}} \right)$ is scored by ${N_{rs}}$ honest readers. Then, the effective score  ${W_{i,j}}$  of review ${R_{i,j}}$ is obtained by first excluding the highest and lowest 10\% scores from  ${N_{rs}}$ readers' scores and then averaging the remaining scores. Finally, we compute the final score ${S_i}$ of paper $i$  according to (3) and (4). The results are shown in Fig. 6 and Fig. 7, where the final scores ${S_i}$  are evaluated with respect to different numbers of malicious nodes  ${N_{mn}}$. We treat the simple average of the review scores, i.e., ${S_i} = \sum\nolimits_j {{Z_{i,j}}} $, as our benchmark. In the simulations, we set $S = 40$, $S' = 80$ , ${W_U} = 90$ and ${N_{rs}} \in \{ 10,{\kern 1pt} {\kern 1pt} {\kern 1pt} {\kern 1pt} 100,{\kern 1pt} {\kern 1pt} {\kern 1pt} 300,{\kern 1pt} {\kern 1pt} {\kern 1pt} {\kern 1pt} 600\} $. Fig. 6 and Fig. 7 show the results for $\sigma _s^2 = 10$, and $\sigma _s^2 = 100$, respectively. As we can see, our scoring method is robust to the attacker’s fake reviews. When more and more malicious nodes are involved (large ${N_{mn}}$), the attacker becomes more successful in biasing the score toward the fake score. Large readership on the PubChain platform means large ${N_{rs}}$ , and large ${N_{rs}}$ makes the system more robust against large ${N_{mn}}$.

\begin{figure}[!t]
	\centering
	\includegraphics[width=3.5in]{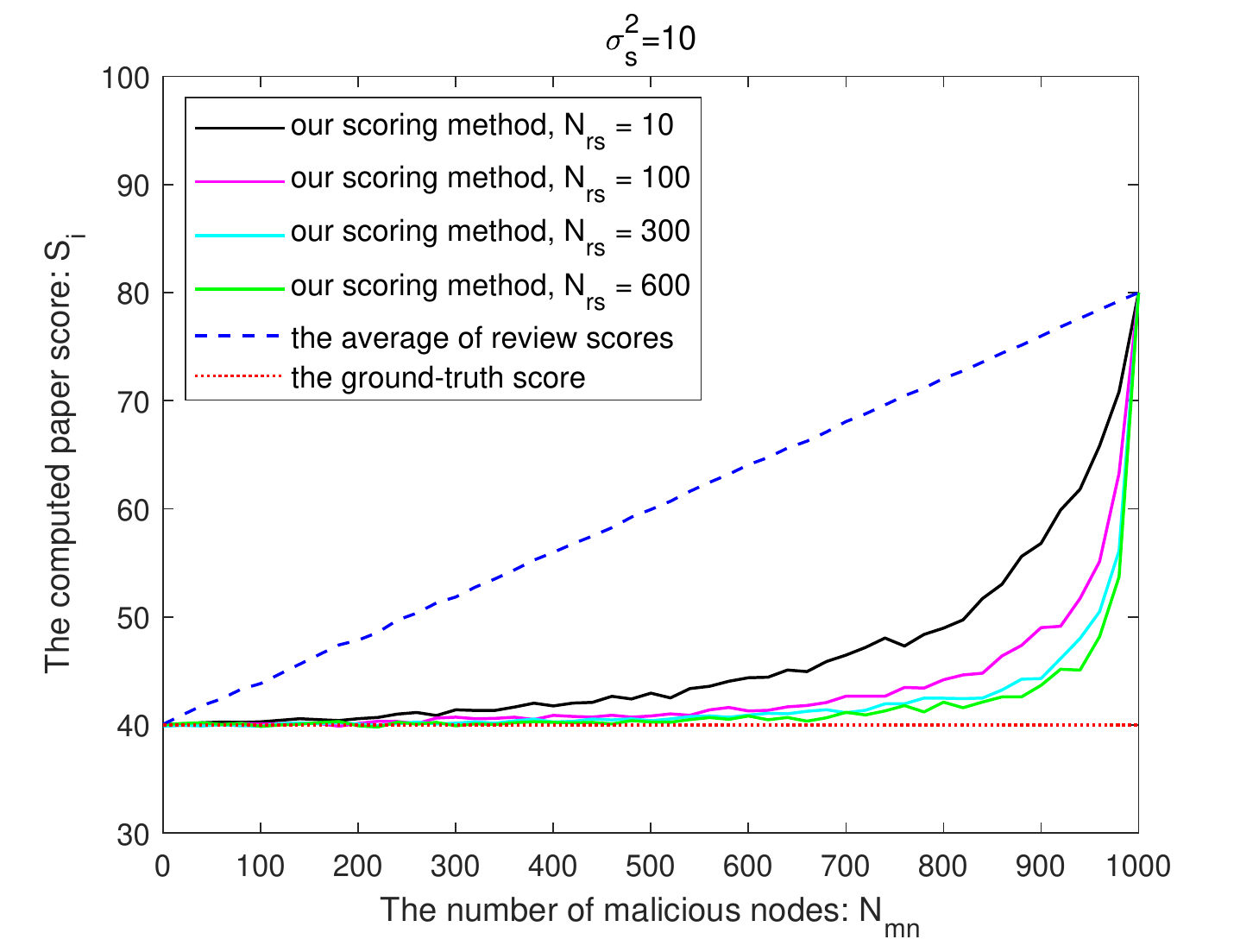}
	\caption{The curve of ${S_i}$ vs. ${N_{mn}}$ with $\sigma _s^2 = 10$ for the first attack strategy.}
\end{figure}

\begin{figure}[!t]
	\centering
	\includegraphics[width=3.5in]{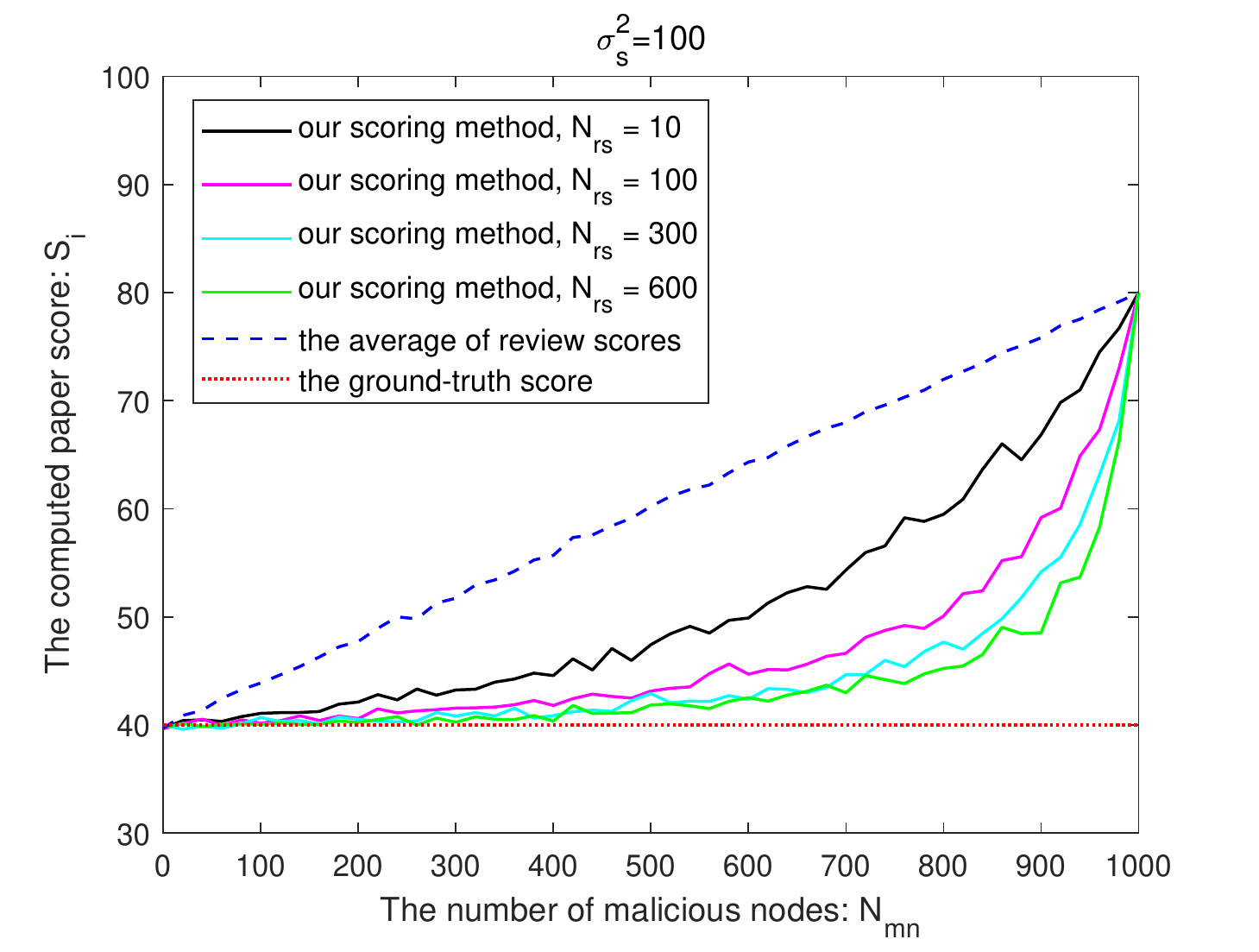}
	\caption{The curve of ${S_i}$ vs. ${N_{mn}}$ with $\sigma _s^2 = 100$ for the first attack strategy.}
\end{figure}

\begin{figure}[!t]
	\centering
	\includegraphics[width=3.5in]{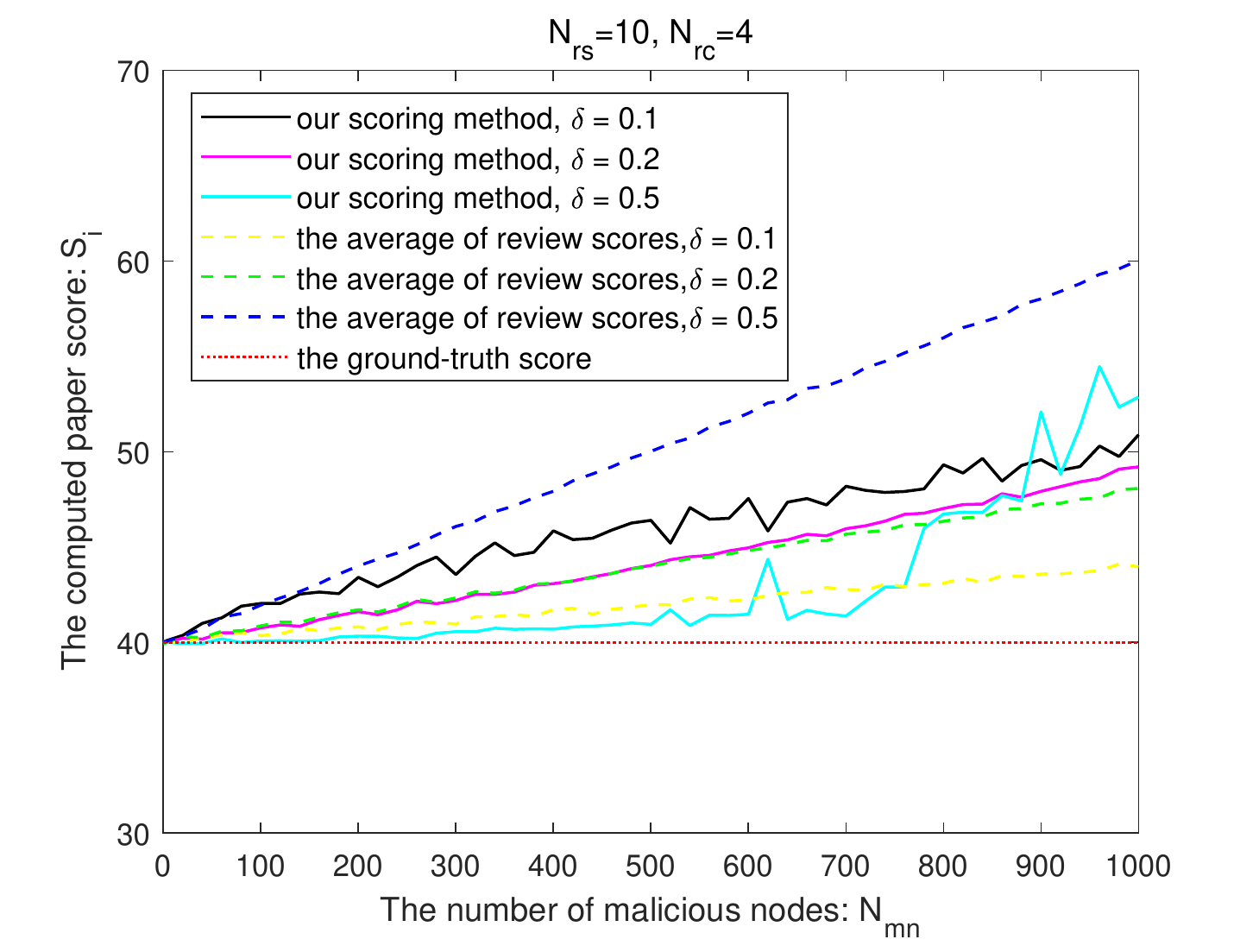}
	\caption{The curve of ${S_i}$ vs. ${N_{mn}}$ with ${N_{rs}} = 10$ for the second attack strategy.}
\end{figure}

\begin{figure}[!t]
	\centering
	\includegraphics[width=3.5in]{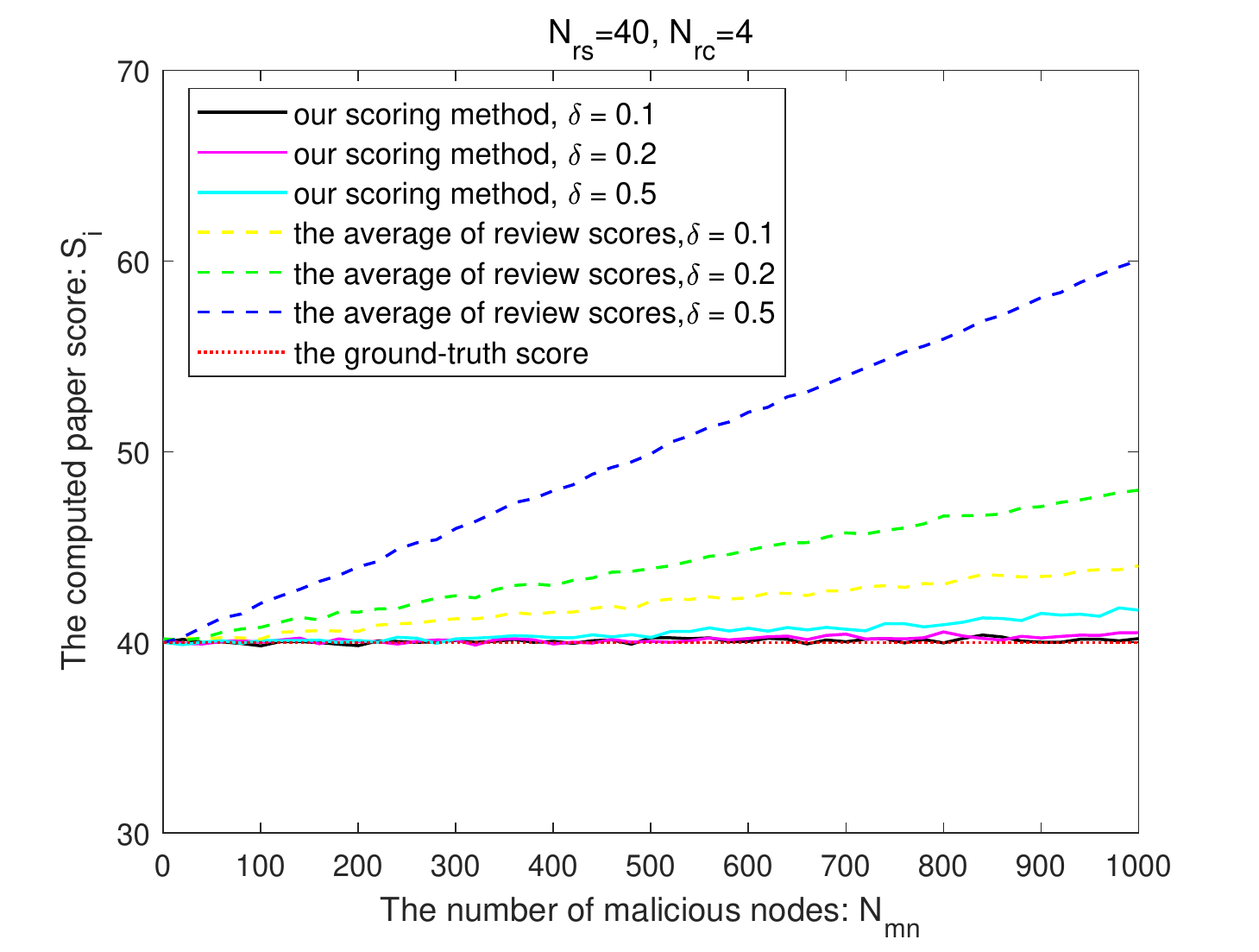}
	\caption{The curve of ${S_i}$ vs. ${N_{mn}}$ with ${N_{rs}} = 40$ for the second attack strategy.}
\end{figure}

The second strategy is to have a fraction $\delta $  of the malicious nodes be fake reviewers and the rest be fake readers; and half of the fake readers will support the fake reviews by giving high scores, and the other half  of the fake readers will attack the honest reviews by giving them low scores. For example, suppose that there are ${N_{mn}} = 100$ malicious nodes and $\delta  = 0.1$. Then, 10 of the malicious nodes are fake reviewers that give review score $S'$ to paper $i$ and 90 malicious nodes are fake readers that can give  a total of ${\rm{9}}0{N_{rc}}$ fake scores to all reviews of paper $i$. Among the ${\rm{9}}0{N_{rc}}$ fake scores to reviews, $45{N_{rc}}$ scores of  ${V_U}$ are given to the fake reviews put up by the attacker (each of the 10 fake reviews is assigned with ${{45{N_{rc}}} \mathord{\left/
		{\vphantom {{45{N_{rc}}} {10}}} \right.
		\kern-\nulldelimiterspace} {10}}$ scores of ${V_U}$), where  ${V_U}$ is a very high score used to support the fake reviews ;  $45{N_{rc}}$ scores of  ${V_L}$ are given to the honest reviews (each honest review is assigned with ${{45{N_{rc}}} \mathord{\left/
		{\vphantom {{45{N_{rc}}} {90}}} \right.
		\kern-\nulldelimiterspace} {90}}$ scores of ${V_L}$), where ${V_L}$ is a very low score used to attack these honest reviews.  

In the simulation, we set $S = 40$, $S' = 80$ , ${W_P} = 90$, ${V_L} = 20$, ${V_U} = 100$, ${N_{rc}} = 4$, $\delta  \in (0.1\,\,,0.2,\,\,0.5\} $.  Fig. 8 and Fig. 9 show the results for ${N_{rs}} = 10$, and ${N_{rs}} = 40$, respectively.  From the results, we can observe that with large ${N_{rs}}$, our scoring method is still robust to this attack strategy.

\section{System Implementation }
We have implemented a proof-of-concept prototype for the PubChain platform. The implementation of the blockchain reuses Ethereum, which means we can realize the virtual machine layer of PubChain using the EVM smart contract mechanism. The prototype uses IPFS for the storage layer. We have deployed the PubChain interface to a network node with address http://120.78.71.240:3000/. Users (i.e., publication players) can use the JSON-RPC protocol to remotely deploy and invoke smart contract via this network node to conduct their activities on PubChain.

\begin{algorithm}[h] 
	\caption{Smart Contract for Posting Papers} 
	\begin{algorithmic}[1] 
		\State {\bf struct} Paper\{
		\State \;\;    {\bf string} ownership;
		\State \;\;    {\bf string} paperName;
		\State \;\;    {\bf string} paperHash;
		\State \;\;    {\bf string} paperKeywords;
		\State \}
		\State {\bf function} stringsEqual(string storage \_a, string memory \_b)  internal view returns ({\bf bool}) \{
		\State {\bf bytes storage} a = bytes(\_a);
		\State {\bf bytes memory b} = bytes(\_b); 
		\State if (a.length != b.length)
		\State \;\; return false;
		\State for (uint i = 0; i < a.length; i ++) \{
		\State \;\; if (a[i] != b[i])
		\State \;\; \;\; return false;
		\State \;\; return true;
		\State \;\; \}
		\State \}
		\State {\bf function} storeData({\bf string} userName\_, {\bf string} paperName\_, {\bf string} paperHash\_, {\bf string} paperKeywords\_) public \{
		\State \;\; require(!stringsEqual(paperInfo[paperName\_].paperName, paperName\_) \&\& !stringsEqual(paperInfo[paperName\_].paperHash, paperHash\_), "one paper one upload");
		\State \;\; paperInfo[paperName\_] = Paper(\{    
		\State \;\;\;\; ownership: userName\_,
		\State \;\;\;\; paperName: paperName\_,
		\State \;\;\;\; paperHash: paperHash\_,
		\State \;\;\;\; paperKeywords:  paperKeywords\_,
		\State \;\; \});
		\State {\bf function} getPaperHash({\bf string} paperName\_) public view returns ({\bf string}) \{
		\State \;\; return paperInfo[paperName\_].paperHash;
		\State \}
		\State  {\bf function} getPaperOwnership(string paperName\_) public view returns ({\bf string}) \{
		\State \;\; return  paperInfo[paperName\_].ownership;
		\State \;\; \}
		\State \}
	\end{algorithmic} 
\end{algorithm}

With smart contracts, we have implemented the functions of paper posting, paper reviewing, review scoring. The script codes of smart contracts are stored on blockchain. The smart contracts are triggered by transactions sent sent to their address on blockchain. Algorithm 1 shows the script codes of the smart contract that implements the function of paper posting.  To post a paper on PubChain, an author needs to carry out the following procedure: 1) upload the paper (possibly including some program codes, multimedia materials) with her/his signature to the IPFS system and obtain the IPFS address of this paper (i.e., the paper hash); 2) include the publication information about the paper, i.e., its ownership (the address of the author on blockchain), IPFS address, paper title, key words, etc.) into a paper metadata record; 3) pack the metadata of the paper to a transaction; 4) issue the transaction to the blockchain system. After the smart contract receives the transaction, it can be executed by some miner to write the metadata of the paper to the blockchain. Fig. 10  shows the window of Remix Ethereum IDE \cite{remix} after the paper posting smart contract is triggered by a transaction that posts our paper to the deployed Ethereum testnet. The procedures and smart contracts for other functions are designed and implemented in similar ways.

Currently, we have not implemented the proposed incentive mechanism that requires extensive modifications on the blockchain program codes.  This is the most important part of our follow-up work.

\begin{figure*}[!t]
	\centering
	\includegraphics[width=6.5in]{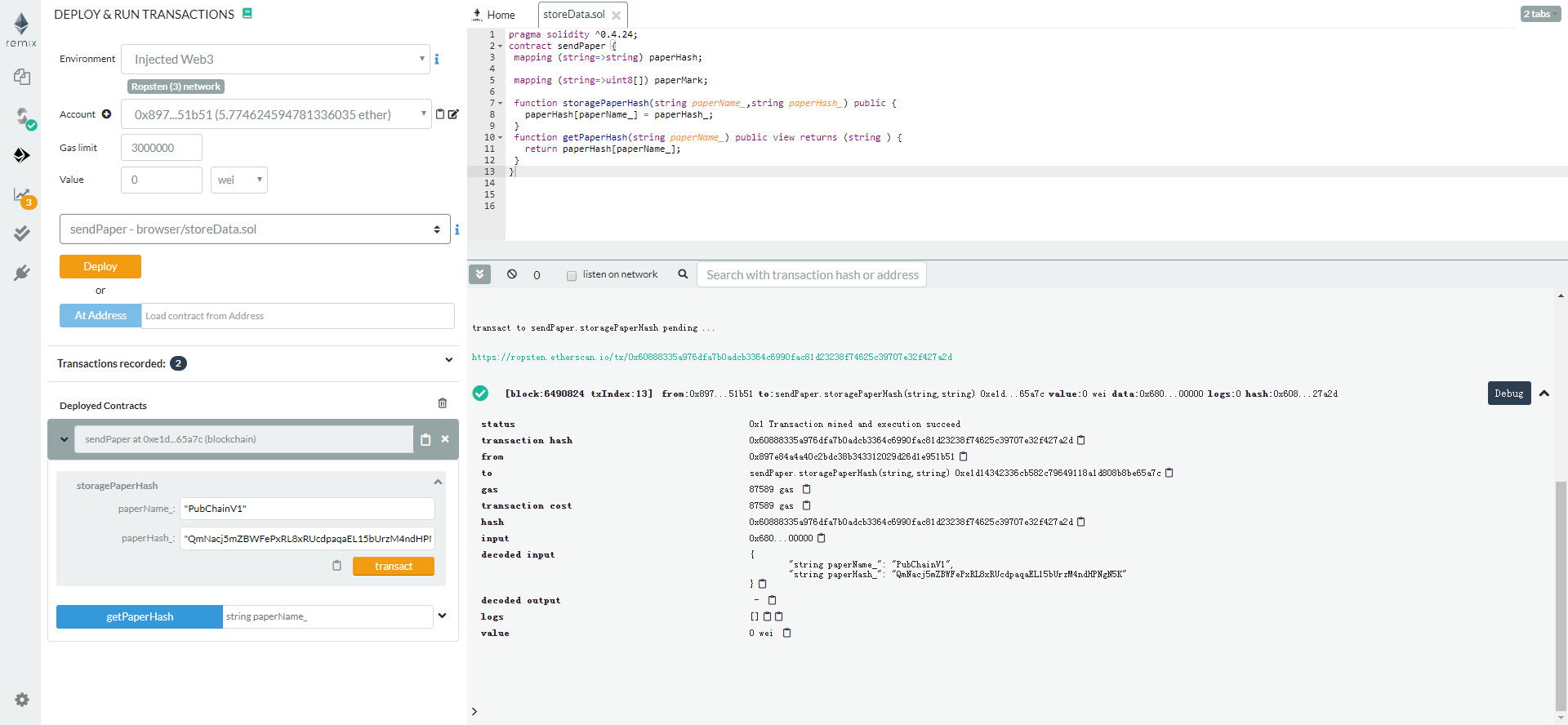}
	\caption{The window of Remix Ethereum IDE after the paper posting smart contract is triggered by a transaction that posts our paper to the Ethereum testnet.}
\end{figure*}

\section{Conclusion}

To overcome the drawbacks and limitations of existing publication platforms for research papers, we exploit recent advances in decentralized technologies (i.e., blockchain, IPFS) to design a decentralized open-access publication platform named  PubChain. Compared with the existing centralized publication platforms, PubChain has several advantages: (i)  PubChain breaks the pay wall imposed by publishers so that everybody can enjoy free access to papers. (ii) PubChain eliminates undesired effects of information islands and has the potential to become a unified database for global sharing and recording of papers. (iii) PubChain, as a decentralized system, provides uninterrupted service without single points of failure. (iv) PubChain incentivizes participants to make positive contributions to the platform with an incentive scheme implemented over blockchain technology.

Importantly, unlike many other publication platforms, PubChain is not meant to be a profit-oriented platform.  The donation of cryptocurrency injects initial financial values to Pubchain. We propose to use a two-way pegging technique to lock donated cryptocurrency to a special address of the parent chain that cannot be spent by any individual address. The project development team, as volunteers, will not receive any cryptocurrency 

This project will be successful only if it can recruit the participation of authors, reviewers, and readers who believe in the tenet of free dissemination and free open access to timely research results.  We invite more volunteers to join the project and work with us to improve the design of Pubchain, and to serve as advocates for the new way of knowledge dissemination for the benefit of humanity.

\bibliographystyle{IEEEtran}

\bibliography{database}

\end{document}